# THIRD-ORDER LAGRANGIAN PERTURBATION THEORY - REALIZATION AT HIGH-SPATIAL RESOLUTION

T. BUCHERT, A. G. WEISS

*Max-Planck-Institut für Astrophysik, 85740 Garching, Munich, Germany*

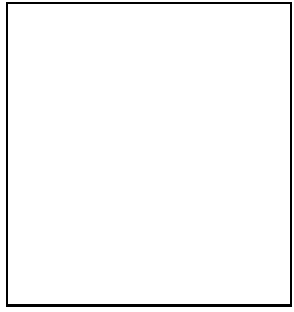

## Abstract

The Lagrangian theory of gravitational instability of homogeneous-isotropic Friedman-Lemaître cosmogonies investigated and solved in the series of papers by Buchert (1989), (1992), Buchert & Ehlers (1993), Buchert (1993a,b), Ehlers & Buchert (1993), is illustrated. The third-order solution of this theory for generic initial conditions is presented and realized in a special case by employing methods of high-spatial resolution of the density field.

## 1  A class of generic third-order solutions

We require that, initially, the peculiar-acceleration $\vec{w}(\vec{X}, t)$ be proportional to the peculiar-velocity $\vec{u}(\vec{X}, t)$:

$$\vec{u}(\vec{X}, t_0) = \vec{w}(\vec{X}, t_0) t_0 \ , \tag{1}$$

where we have defined the fields as usual (compare Peebles 1980, Buchert 1992). Henceforth, we denote the peculiar-velocity potential at the initial time $t_0$ by $\mathcal{S}$, $\vec{u}(\vec{X}, t_0) =: \nabla_0 \mathcal{S}$, and the peculiar-gravitational potential at $t_0$ by $\phi$, $\vec{w}(\vec{X}, t_0) =: -\nabla_0 \phi$, where $\nabla_0$ denotes the nabla operator with respect to Lagrangian coordinates $\vec{X}$. This restriction has proved to be appropriate for the purpose of modeling large-scale structure, since, for irrotational flows, the peculiar-velocity field tends to be parallel to the peculiar-acceleration after some time, related to the existence of growing and decaying solutions in the linear regime. Still, *any* generic function ($\mathcal{S}$ or $\phi$) can be given initially.

With a superposition ansatz for Lagrangian perturbations of a flat homogeneous and isotropic background the following family of trajectories $\vec{x} = \vec{f}(\vec{X}, a)$ as *irrotational* solution of the Euler-Poisson system up to the third order in the perturbations from homogeneity has been obtained (Buchert 1993b). The general set of initial conditions $(\phi(\vec{X}), \mathcal{S}(\vec{X}))$ is restricted according to $\mathcal{S} = -\phi\, t_0$ (see equation (1)). (The parameter $\varepsilon$ is thought to be absorbed into the amplitudes of the perturbation potentials; $a(t) = (t/t_0)^{2/3}$):

$$\vec{f} = a\,\vec{X} + q_z(a)\,\nabla_0 \mathcal{S}^{(1)}(\vec{X}) + q_{zz}(a)\,\nabla_0 \mathcal{S}^{(2)}(\vec{X})$$
$$+ q^a_{zzz}(a)\,\nabla_0 \mathcal{S}^{(3a)}(\vec{X}) + q^b_{zzz}(a)\,\nabla_0 \mathcal{S}^{(3b)}(\vec{X}) - q^c_{zzz}(a)\,\nabla_0 \times \vec{\mathcal{S}}^{(3c)}(\vec{X}) \qquad (2)$$

with:

$$q_z = \left(\frac{3}{2}\right)(a^2 - a)$$

$$q_{zz} = \left(\frac{3}{2}\right)^2 \left(-\frac{3}{14}a^3 + \frac{3}{5}a^2 - \frac{1}{2}a + \frac{4}{35}a^{-\frac{1}{2}}\right)$$

$$q^a_{zzz} = \left(\frac{3}{2}\right)^3 \left(-\frac{1}{9}a^4 + \frac{3}{7}a^3 - \frac{3}{5}a^2 + \frac{1}{3}a - \frac{16}{315}a^{-\frac{1}{2}}\right)$$

$$q^b_{zzz} = \left(\frac{3}{2}\right)^3 \left(\frac{5}{42}a^4 - \frac{33}{70}a^3 + \frac{7}{10}a^2 - \frac{1}{2}a + \frac{4}{35}a^{\frac{1}{2}} + \frac{4}{105}a^{-\frac{1}{2}}\right)$$

$$q^c_{zzz} = \left(\frac{3}{2}\right)^3 \left(\frac{1}{14}a^4 - \frac{3}{14}a^3 + \frac{1}{10}a^2 + \frac{1}{2}a - \frac{4}{7}a^{\frac{1}{2}} + \frac{4}{35}a^{-\frac{1}{2}}\right)$$

where the initial displacement vectors have to be constructed by iteratively solving the 7 Poisson equations:

$$\Delta_0 \mathcal{S}^{(1)} = I(\mathcal{S}_{,i,k})\,t_0$$
$$\Delta_0 \mathcal{S}^{(2)} = 2II(\mathcal{S}^{(1)}_{,i,k})$$
$$\Delta_0 \mathcal{S}^{(3a)} = 3III(\mathcal{S}^{(1)}_{,i,k})$$
$$\Delta_0 \mathcal{S}^{(3b)} = \sum_{a,b,c} \epsilon_{abc} \frac{\partial(\mathcal{S}^{(2)}_{,a}, \mathcal{S}^{(1)}_{,b}, X_c)}{\partial(X_1, X_2, X_3)}$$
$$(\Delta_0 \vec{\mathcal{S}}^{(3c)})_k = \epsilon_{pq[j} \frac{\partial(\mathcal{S}^{(2)}_{,i]}, \mathcal{S}^{(1)}_{,p}, X_q)}{\partial(X_1, X_2, X_3)}$$

($i, j, k = 1, 2, 3$ cyclic ordering).

The scalar potential $\mathcal{S}^{(3b)}$ and the vector potential $\vec{\mathcal{S}}^{(3c)}$ generate interaction among the first- and second-order perturbations. Note that the general interaction term is not purely longitudinal. Inspite of the irrotationality of the flow in Eulerian space, vorticity is generated in Lagrangian space starting at the third order.

An important remark relevant to any realization of the solution (2) concerns the possibility of setting $\mathcal{S}^{(1)} = \mathcal{S} t_0$ without loss of generality, if periodic boundary conditions are imposed on $\mathcal{S}$ (compare Buchert 1992, Ehlers & Buchert 1993). In the latter paper it is proved that the solution (2) is unique for such boundary conditions. With this setting, the first-order solution reduces to the well-known "Zel'dovich approximation" (1970) as the only model which depends *locally* on the initial conditions.

The presented model is currently being compared with numerical simulations. We analyze cross-correlations for various power spectra with PM simulations (Buchert *et al.* 1993 a,b), and special cluster models at high resolution using a tree-code (Buchert *et al.* 1993 c).

Parallel attempts to investigate Lagrangian perturbation solutions are made by Moutarde *et al.* (1991), Bouchet *et al.* (1992) and Lachièze-Rey (1993).

## 2  Realization at high-spatial resolution

Let us use for the illustration a coherent random fluctuation field as initial condition. We specify this to standard HDM initial conditions in a box of size 50Mpc h$^{-1}$ normalized to $\sigma_8 = 1$ using top-hat smoothing.

The solution (2) is realized for the first- and third-order approximations at this amplitude (see *Figs.1*). Since the collapse process is significantly accelerated by the higher-order corrections, the first-order approximation represents an earlier stage in the evolution of the clustering process.

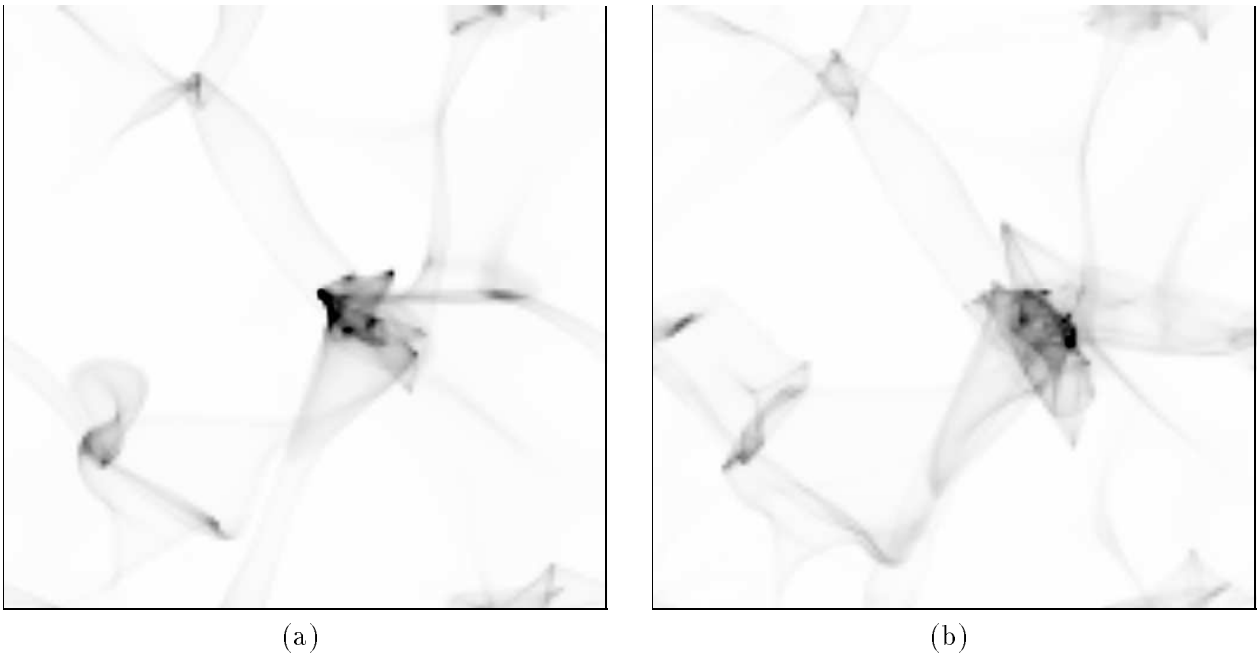

(a) (b)

Figure 1: A thin slice through a 3D HDM realization at high-spatial resolution ($1024^3$ trajectories collected into a $128^3$ pixel grid) is shown for the first-(a) and the third-order approximation (b) for the evolution stage $z = 0$ for the normalization $\sigma_8 = 1$.

The models presented in *Fig.1* have been realized at high-spatial resolution employing the methods described in Buchert & Bartelmann (1991). In particular, we calculate the initial condition at $128^3$ grid points and interpolate 12 times trilinearly between the grid points to obtain an effective resolution of $1024^3$ particles. This interpolation yields highly accurate results, since the initial fluctuation field is sufficiently smooth and the mapping is analytical providing formally infinite resolution.

**Acknowledgements.** We are greatful to Ed Bertschinger and Rien van de Weygaert for providing us their initial conditions code.